\documentclass[preprint,12pt]{elsarticle}
\usepackage[utf8]{inputenc}
\usepackage[english]{babel}

\usepackage{csvsimple}
\usepackage{graphicx}
\usepackage{enumerate}
\usepackage{subcaption}
\usepackage{lineno,hyperref}
\graphicspath{{./Figs/}{./}}
\usepackage{amsfonts,amssymb,amsmath,amscd,amsthm,latexsym,color}
\usepackage[svgnames]{xcolor}
\usepackage{geometry}
\usepackage{tabularx,url}
\usepackage[textwidth=8em,textsize=small]{todonotes}
\usepackage{tikz}
\usetikzlibrary{arrows,shapes,positioning,shadows,trees}

\tikzset{
  basic/.style  = {draw, text width=5cm, drop shadow, font=\sffamily, rectangle},
  root/.style   = {basic, rounded corners=2pt, thin, align=center,
                   fill=green!30},
  level 2/.style = {basic, rounded corners=6pt, thin,align=center, fill=green!60,
                   text width=8em},
  level 3/.style = {basic, thin, align=left, fill=pink!60, text width=3.5cm}
}
\journal{Environmental Modelling and Software}

\biboptions{authoryear}

\newcommand{\new}[1]{\textcolor{black}{#1}}
\newcommand{\mnew}[1]{\textcolor{black}{#1}} 

\begin{document}

\begin{frontmatter}

\title{Sensitivity analysis of spatio-temporal  models describing nitrogen transfers, transformations and losses at the  landscape scale}

\author[1]{Jordi Ferrer Savall}
\author[1]{Damien Franqueville}
\author[2]{Pierre Barbillon\footnote{Corresponding author: {pierre.barbillon@agroparistech.fr}}}
\author[1]{Cyril Benhamou}
\author[3]{Patrick Durand}
\author[4]{Marie-Luce Taupin}
\author[2]{Hervé Monod}
\author[1]{Jean-Louis Drouet}

\address[1]{UMR ECOSYS, INRA, AgroParisTech, Université Paris-Saclay, 78850, Grignon, France.}
\address[2]{UMR MIA-Paris, AgroParisTech, INRA, Université Paris-Saclay, 75005, Paris, France.}
\address[3]{UMR SAS, INRA,  Agrocampus Ouest. 84215, Rennes, France.}
\address[4]{UMR MaIAGE, INRA, Université Paris-Saclay, 78350  Jouy-en-Josas, France.}

\begin{abstract}
Modelling complex systems such as agroecosystems often requires the quantification of a large number of input factors. Sensitivity analyses are useful to \mnew{determine} the appropriate spatial and temporal resolution of models and to reduce the number of factors to be measured or estimated accurately. Comprehensive spatial and \mnew{temporal}
sensitivity analyses were applied to the NitroScape model, a deterministic spatially distributed model describing nitrogen transfers and transformations in rural landscapes. Simulations were led on a theoretical landscape that represented five years of intensive farm management and covering an area of $3\, km^2$. Cluster analyses were applied to summarize
the results of the sensitivity analysis on the ensemble of model outputs. 
The methodology we applied is useful to synthesize sensitivity analyses of models with multiple space-time input and output variables and could be ported to other models than NitroScape.
\end{abstract}


\end{frontmatter}


\section{Introduction}

A main agro-environmental and socio-economic challenge of sustainable agriculture is to maintain agricultural production while reducing the use of nitrogen inputs.
The generalized use of artificial nitrogen fertilizers feeds a cascade of processes that releases nitrogen surplus to the local environment and pollutes the air, soils and waterways.
Losses of reactive forms of nitrogen ($N_r$) have an overall negative impact on ecosystems, economy and human health. \mnew{They} cause eutrophication, biodiversity loss, soil acidification and degradation of drinking water sources 
\citep{galloway2003nitrogen}.

A better understanding of the nitrogen cascade in agroecosystems is required to find innovative ways to reduce losses at each step of the cascade. 
To this end, mathematical models have been developed, evaluated and applied to quantitatively describe nitrogen transfers and transformations at various spatio-temporal scales.
Agro-environmental models are often complex, describing a broad array of phenomena (physical processes, bio-transformations and farm functioning), and using a large number of inputs (model parameters, initial conditions and continuously-fed data on meteorology, soil properties, field management). 
\new{Estimating accurately these inputs often requires a large amount of data. Moreover, collecting this data on field is time-consuming and costly \citep{drouet2011sensitivity}.}

Therefore, determining the \new{spatial (horizontal and vertical) and temporal} resolutions at which model inputs should be measured or estimated is a matter of great practical importance both for the statistical interpretation of field data, and for the meaningful communication of model predictions. Likewise, since the precision of simulations with respect to space and time may influence the model outputs, the optimal spatial granularity and temporal accuracy at which simulations should be run has to be determined prior to using any model to assess mitigation options on real systems. Hence, the effect of the spatial and temporal resolution of simulations should be evaluated together with the effect of uncertainty in model inputs, and their effects on model outputs should be quantitatively compared with each other \cite{bishop2006geostatistical}.

\mnew{Until now}, a wide variety of techniques have been developed to perform sensitivity analysis in spatially and dynamic models at different stages \new{\citep{faivre2013analyse,ghanem2017handbook}}. Methods for exploring model inputs may range from the simplest one-factor-at-a-time screening techniques proposed by \citet{morris1991factorial} to complete factorial designs  \new{via fractional factorial designs
\citep{chen2011fractional} and space-filling designs \citep{damblin2013numerical}}. \new{Sensitivity
analyses based on variance may be performed at each temporal step or at given spatial locations 
and the spatial distribution of the indices, for instance, may be of interest to analyze \citep{marrel2011global}.
Otherwise, sensitivity analyses can also be performed
on a spatial or temporal aggregation of outputs \citep{moreau2013approach}.}
Furthermore, outputs may be aggregated at 
different scales of description \citep{ligmann2013spatially}.
\new{When a model suffers from a deep uncertainty, several scenarios are considered to describe plausible futures. \citet{gao2016robust} and
\citet{gao2016incorporating} proposed an extension of the variance-based sensitivity analysis method and of the Morris method for conducting robust sensitivity
analyses under deep uncertainty.}
\new{Some extensions of sensitivity indices designed for scalar outputs were proposed recently in the literature.
\citet{gamboa2014sensitivity} provided a generalization of the Sobol' indices \citep{saltelli2000sensitivity}
to multidimensional and functional outputs. \citet{de2017sensitivity} proposed an extension of the Hilbert-Schmidt Dependence
Criterion (HSIC) \citep{gretton2005measuring,da2015global} for a spatio-temporal model.
}

The purpose of this paper is to provide some novel analytical and visualization methods to carry out a comprehensive evaluation of the effect of a set of defined input factors on a set of spatially distributed model outputs.
A central concern of the current study is to put forward some tools that allow integrating the results of several sensitivity analyses carried on multiple model outputs into 
\new{summarized indicators.}
\new{These tools could be used as very first exploration of a complex model ($i.e.$ with numerous inputs, outputs and biophysical processes) by detecting which inputs affect the outputs and by grouping outputs which are mainly affected by the same sets of inputs. 
Moreover, the visualization methods facilitate spatial and temporal sensitivity analyses by representing synthetically the behavior of the model with respect to its inputs along time or across space.    
}

To this end, we used the study case of a small theoretical landscape of a few tens of square meters, on which we applied a model describing the cascade of the reactive forms of nitrogen ($N_r$) in landscapes. The sensitivity analysis of the model was carried out by evaluating the effects of various types of input factors ($i.e.$ the spatial resolution of the model, biophysical features of the landscape, agricultural management practices) on several spatially-distributed outputs describing the nitrogen cascade and $N_r$ losses in the environment ($e.g.$ soil ammonium and nitrate amounts and concentrations, emissions of nitrogen oxides and ammonia from the soil to the air, ammonium and nitrate discharge at the catchment outlet).

\section{Materials and methods}

\subsection{Description of the biophysical model and the study case}

\begin{figure}
\begin{subfigure}{\textwidth}
  \centering
  \includegraphics[width=0.9\linewidth]{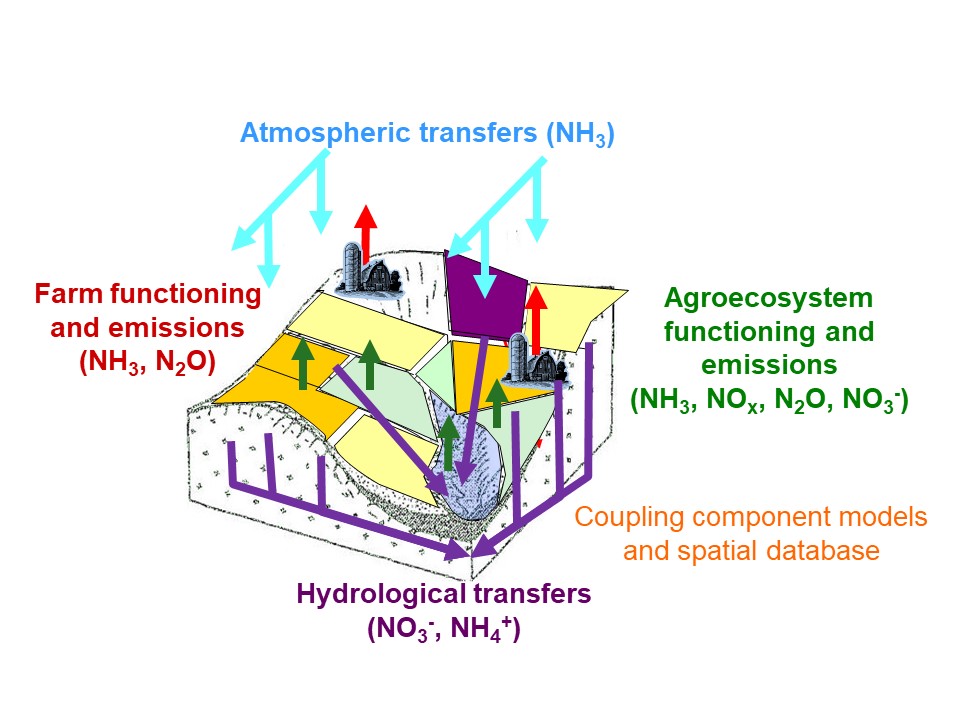}
 \caption{}
  \label{fig:Model}
\end{subfigure} \\
\begin{subfigure}{\textwidth}
  \centering
  \begin{tabular}{cc}
\includegraphics[width=0.5\linewidth]{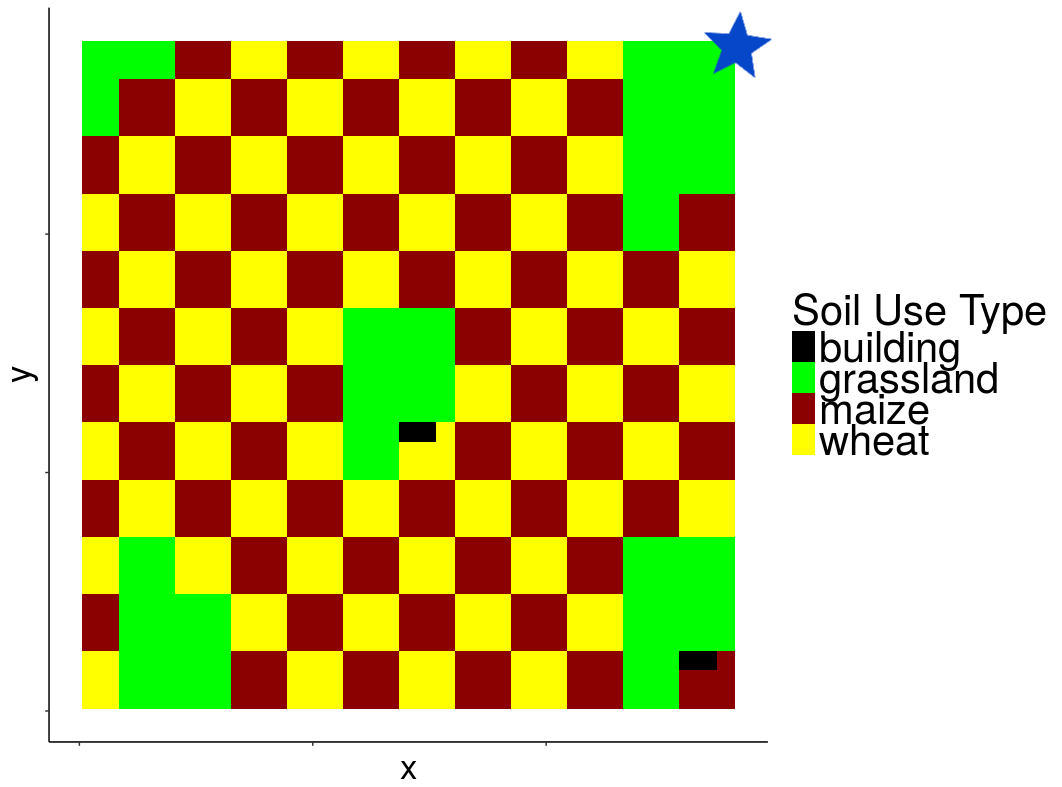}
 &
  \includegraphics[width=0.5\linewidth]{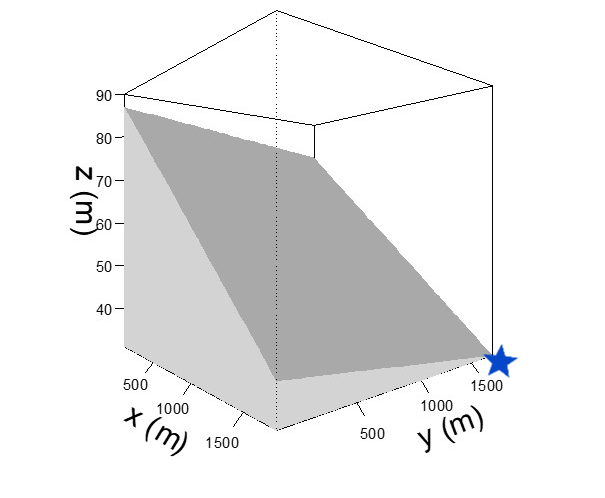} 
  \end{tabular}
 \caption{}
  \label{fig:Landscape}
\end{subfigure}
\caption{Scheme of the NitroScape model (a). Land use and topography of the theoretical landscape (b), shown here for the highest spatial horizontal resolution of the model (grid cells of size 12.5 m x 12.5 m each). The blue star indicates the catchment outlet.}
\label{fig:Nitroscape}
\end{figure}

NitroScape is a deterministic, spatially distributed and dynamic model describing $N_r$ transfers and transformations in rural landscapes \citep{duretz2011nitroscape}. It couples four modules characterizing farm management, biotransformations and transfers by the atmospheric and the hydrological pathways (Fig. \ref{fig:Model}). It simulates the concentrations and fluxes, including the losses, of different forms of $N_r$ (reduced forms (ammonia $NH_3$, ammonium $NH_4^+$), inorganic oxidized forms (nitrate $NO_3^-$, nitrogen oxides $NO_x$ and nitrous oxide $N_2O$) and organic forms (manure, crop residues) within and between several landscape compartments: the atmosphere, the hydro-pedosphere (soil, water table, groundwater and streams) and the terrestrial agroecosystems (livestock buildings, croplands, grasslands and semi-natural areas). 

\paragraph{Software availability}
~ \\
\begin{tabularx}{\textwidth}{lX}
Name of software:& NitroScape\\
Developer:& Jean-Louis Drouet, Camille Chambon\\
Contact:& UMR 1402 ECOSYS, route de la ferme, 78850 Thiverval-Grignon; tel: +33 1 30 81 55 68; fax: +33 1 30 81 55 63; email: jean-louis.drouet@inra.fr\\
Year first available:& 2011\\
Hardware required:& PC (or cluster of PCs to reduce computing time) with Unix (preferably Linux Fedora)\\
Software required:& OpenPALM coupler (\url{http://www.cerfacs.fr/globc/PALM\_WEB/}), component models: CERES-EGC, FARM-EF, FIDES-3D-SURFATM, TNT\\
Program langages:& fortran, C, C++, java, R\\
Program size:& several thousands of lines\\
Software availability:& source code can be provided through collaborative arrangements\\
Cost:& free through collaborative arrangements\\ \\
\end{tabularx}

NitroScape was applied to simulate the nitrogen cascade and $N_r$ losses on a simplified theoretical landscape (Fig. \ref{fig:Landscape}) of 300 ha corresponding to an intensive rural area with a succession of maize and wheat crops in a checkerboard distribution (125 ha each crop), pig farming buildings (two separate buildings, one ha each) and unmanaged grasslands (five plots scattered within the landscape and comprising 48 ha in total). Each square of the checkerboard was a set of grid cells whose size corresponded to the horizontal spatial resolution of the model. For instance, when the horizontal resolution of the model was 25 m x 25 m, the landscape was represented as a checkerboard of 10 x 10 squares, each square being represented by 7 x 7 grid cells. Topography was characterized by a linear slope with a gradient of 50 m between the highest and the lowest parts of the landscape. Meteorology was characterized by humid climatic conditions and little temperature contrasts. Meteorological data used for simulations were measured with a meteorological station located on the Kervidy-Naizin catchment (Brittany, $48^{\circ}01$’N, $2^{\circ}83$'O) between 2007 and 2011. Atmospheric dispersion, transfer and deposition were not taken into account in this exploratory study since running the atmospheric component of NitroScape is very time-consuming. Further specifications on the NitroScape model and the theoretical landscape can be found in \citet{duretz2011nitroscape}. 

Simulations were performed at a daily time step and integrated over a five-year period, starting from January 1$^{st}$, 2007. The first two years of simulation were used for model initialization and the sensitivity analysis used the results provided by the last three years of simulation only. Daily outputs were sampled from the variables simulated at the catchment outlet and monthly outputs were sampled from results obtained at different locations within the landscape.

\subsection{Analyzing a spatio-temporal model}

\begin{figure}
\centering
\begin{tikzpicture}[
  level 1/.style={sibling distance=50mm},
  edge from parent/.style={->,draw},
  >=latex]

\node[root] {Design of experiments and runs of NitroScape}
  child {node[level 2] (c1) {Temporal analyses}}
  child {node[level 2] (c2) {Spatial analyses}}
  child {node[level 2] (c3) {Global analyses}};

\begin{scope}[every node/.style={level 3}]
\node [below of = c1, xshift=25pt] (c11) {Spatial aggregation \\ and outflows};
\node [below of = c11,yshift=-15pt] (c12) {Visualization \\ (central map and \\ middle region)};
\node [below of = c12,yshift=-10pt] (c13) {KML clustering};
\node [below of = c13,yshift=5pt] (c14) {SA at each time};
\node [below of = c14,yshift=-4pt] (c15) {PCA with \\ 3 components};
\node [below of = c15,yshift=-10pt] (c16) {SA on each PCA component};

\node [below of = c2, xshift=25pt] (c21) {Temporal \\ aggregation};
\node [below of = c21,yshift=-15pt] (c22) {Visualization \\ (central map and \\ standard deviation)};
\node [below of = c22,yshift=-10pt] (c23) {SA on each grid cell};
\node [below of = c23] (c24) {PCA with \\ 3 components};
\node [below of = c24,yshift=-10pt] (c25) {SA on each PCA component};

\node [below of = c3, xshift=25pt] (c31) {Temporal and \\ spatial aggregation};
\node [below of = c31,yshift=-10pt] (c32) {Computation of \\ sensitivity indices};
\node [below of = c32,yshift=-16pt] (c33) {Hierarchical \\ clustering on \\ sensitivity indices};
\node [below of = c33,yshift=-10pt] (c34) {PCA visualization};
\end{scope}

\foreach \value in {1,...,6}
  \draw[->] (c1.195) |- (c1\value.west);

\foreach \value in {1,...,5}
  \draw[->] (c2.195) |- (c2\value.west);

\foreach \value in {1,...,4}
  \draw[->] (c3.195) |- (c3\value.west);
\end{tikzpicture}
\caption{Workflow of sensitivity analyses (see Subsection 2.2 for details). SA means Sensitivity analysis and PCA means Principal component analysis.}
\label{fig:workflow}
\end{figure}
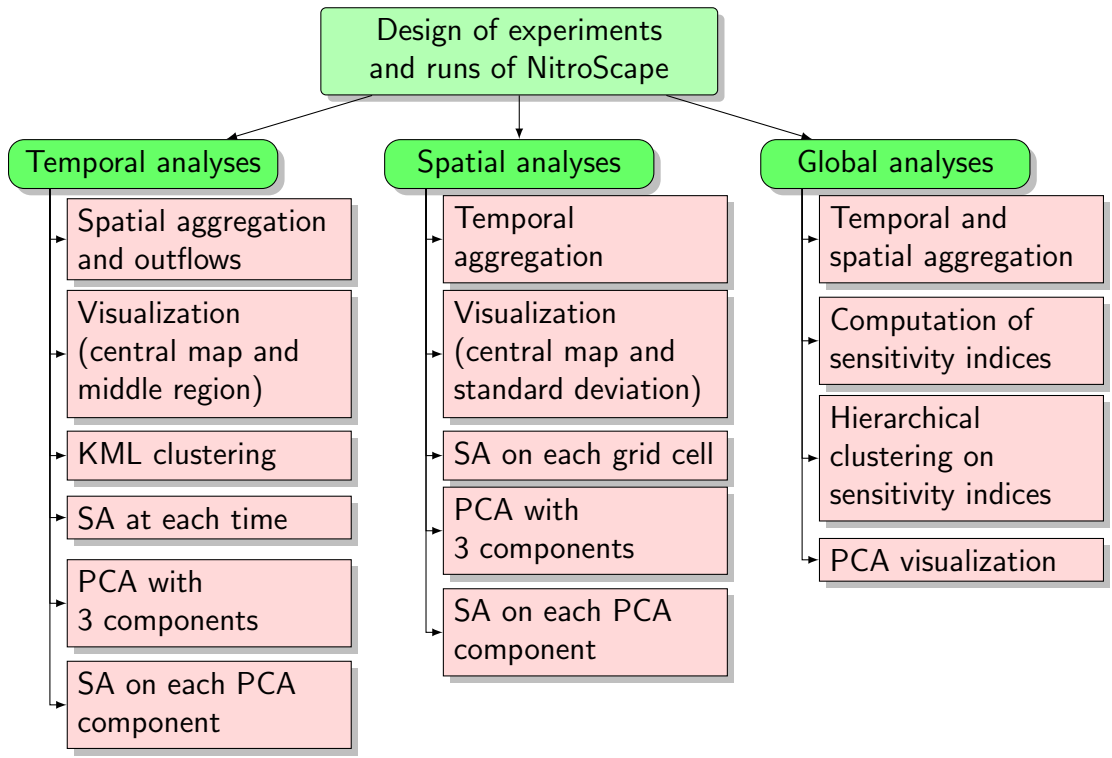

\new{The workflow used to analyze the NitroScape model is described in Figure \ref{fig:workflow}. Three levels of analyses are considered: a temporal analysis where the outputs are spatially-aggregated, a spatial analysis where the outputs are temporally-aggregated and a global analysis in which aggregation is both spatial and temporal. Details on the different methods of analysis are provided hereafter.}

\subsubsection{Design of numerical experiments}

\new{Eleven input factors were selected to evaluate the sensitivity of model outputs to model inputs (Tab. \ref{tab : factors}). 
We chose those factors because they represent the three main types of input factors used by the spatial,
dynamic and integrated NitroScape model: the spatial ($i.e.$ horizontal and vertical) resolution of the model
(quantitative input factors A and B), the biophysical parameters which affect ${a \ priori}$ the ${N_r}$ 
fluxes in the agro-pedo-hydrosphere (quantitative input factors C to I) and two farm practices which mainly affect ${N_r}$
fluxes and concentrations (qualitative and quantitative input factors J and K).}
\new{For this exploratory study, the effect of the spatial organization of the landscape on ${N_r}$ fluxes, including losses, was evaluated through the single arrangement of fields and farm buildings set in the theoretical landscape.}
\new{Hereafter, no cross correlations between factors were considered since we aimed at identifying the plain effects of
each factor and avoiding confusion. Moreover, no spatial correlation was modeled for input factors since they were set as constant over the whole landscape.}
\\ 

\begin{table}[ht!]
     \centering
       \begin{tabular}{|c|c|c|c|}%
    \hline
    Factor&Description&Levels&Unit \\
    \hline
    A&Grid cell width (horizontal resolution)&{12.5, 25, 50}&m\\
    B&Soil layer depth (vertical resolution)&{0.02, 0.05, 0.1}&m\\
    C&Soil lateral transmissivity&{2, 8, 15}&$m^2/$day \\
    D&Depth of exponential decrease of soil transmissivity&{0.001, 0.01, 0.1}&m\\
    E&Surface layer (HS) depth&{0.2, 0.3, 0.4}&m\\
    F&Soil porosity of the surface layer&{0.12, 0.24, 0.48}&-\\
    G&Ratio of soil microporosity to macroporosity&{0.5, 1, 1.2}&-\\
    H&Intermediate layer (HI) depth&{0.6, 0.9, 1.2}&m\\
    I&Ratio of microporosity between layers (HI/HS)&{1, 0.75, 0.5}&-\\
    J&Type of nitrogen fertilization&{OL, OF, INO}&-\\
    K&Amount of nitrogen in fertilizer&$X \pm 20\%$&kg N ha$^{-1}$\\
    \hline
  \end{tabular}
  \caption{Input factors of NitroScape that were varied in the numerical experiments. OL: organic liquid manure, OF: organic solid fertilizer, INO: inorganic mineral fertilizer. The amounts of nitrogen in fertilizer were set at three levels: a fixed value (X) that depends on the type of fertilization, the number of applications and the type of crops (average value: $180 \ kg \ N \ ha^{-1} \ yr^{-1}$), and two values at $\pm 20\%$ of the fixed value. \new{Input factors from  C to I were set as constant throughout the landscape and input factors J and K took non-zero values for fertilization events only.}}
  \label{tab : factors}
\end{table}

The effect of model inputs was evaluated on \new{all the} 29 $N_r$-related model outputs: 5 variables describing the fluxes and concentrations of $N_r$ at the catchment outlet ($e.g.$ daily $NO_3^-$ concentration and amount), 9 spatially-distributed variables describing the fluxes at the interface between compartments ($e.g.$ evapotranspiration, amount of mineralized $NH_4^+$ or $NO_3^-$) and 15 spatially-distributed variables describing the local state of the compartments ($e.g.$ $NH_4^+$ or $NO_3^-$ concentration in groundwater or in soil).

\new{Given the size of the numerical experiment and the mixture of quantitative and qualitative factors, we adopted a screening-design approach using} a fractional factorial design (FFD) \citep{saltelli2000sensitivity} of size 243, that corresponds to 243 configurations combining the 11 input factors, each with 3 levels. \new{The discretization into three levels for quantitative factors enables the detection of non monotonic effects.}
The design was generated using the R package \textit{Planor} \citep{kobilinsky2012planor}.
\new{The resulting FFD was obtained from a design of resolution 5, which means that this design makes it possible to determine for each output the main effects and the pairwise interactions of input factors, without confounding effect between factors \citep{box1987empirical}, in a model of analysis of variance (ANOVA). This design was also saturated since there was no residual degree of freedom to estimate the variance.}

%

\subsubsection{Aggregation of simulated outputs}

Spatially-distributed outputs formed large sets of data that were difficult to handle with conventional statistical tools: each output was described by a matrix of 243 rows and up to more than $7.10^5$ columns. Each row corresponded to each configuration of the FFD and each column corresponded to each output variable in each grid cell of the theoretical landscape. For instance, for the highest horizontal resolution ($i.e.$ grid cells of size 12.5 m x 12.5 m each, Fig. 1b), the theoretical landscape included 19,600 grid cells, each characterized by the value of the 36 simulated monthly output variables, which resulted in 705,600 columns. For this reason, the output variables were spatially- or temporally-aggregated to produce different types of data sets: time series describing spatially-aggregated outputs were used to perform temporal sensitivity analysis (Section 3.1), while maps of temporally-aggregated outputs were used for spatial sensitivity analysis (Section 3.2). All output variables were also spatially- and temporally-aggregated to provide a synthetic view of the sensitivity of model outputs to input factors (Section 3.3).

\subsection{Visualization}
\label{ssec:visualization}

\new{
The time series (resp. the map) of the highest densities of outputs were plotted to summarize of the temporal (resp. spatial) outputs. They were obtained from a functional boxplot of the highest density region (HDR) \citep{hyndman2010rainbow}. 
HDR boxplots were defined by computing a bivariate kernel density estimate on the first two principal components of a principal component analysis performed on the time-series (resp. maps), and then applying the bivariate HDR boxplot of \citet{hyndman1996computing}. 
The central time series or map made more physical meaning than a pointwise average. Regarding time series, 
the middle region that contained half of the time series was also plotted. We adapted some functions of the \textit{Rainbow} R-package to obtain these plots.
}
\\

\subsection{Sensitivity analysis}

%

\new{Sensitivity analyses were performed on the basis of an ANOVA model. The R package \textit{Multisensi} \citep{lamboni2011multivariate} was used.
For each configuration $i$ of the FFD ($i=1,\ldots, n$; $n=243$), let $Y_i$ be the outputs of interest ($Y_i = f(x_{i,1},\ldots,x_{i,p})$; $p=11$; factor number $j=1,\ldots,p$ corresponding to letters A,\ldots,K respectively). These outputs can be spatially- or temporally-aggregated or be the projection of the time series or the spatial map of a given output on one of the three axes of the PCA (see Section 2.5). The notation $x_{i,j}$ stands for the input factor $j$ of the configuration $i$ of the FFD. The three different levels of each factor $j$ are denoted by $k$ ($k=1,2,3$).
An ANOVA model was adjusted to analyze main effects and second order interactions between factors:
\begin{equation}
\nonumber
 Y_i = f(x_{i,1},\ldots,x_{i,p}) = \mu + \sum_{j=1}^p \alpha^{(j)}_{x_{i,j}} + \sum_{1\le j<j' \le p} \beta^{(j,j')}_{x_{i,j},x_{i,j'}} + E_i  
\end{equation}
where $\alpha^{(j)}_{x_{i,j}}$ is the main effect of factor $j$ on the output and $\beta^{(j,j')}_{x_{i,j},x_{i,j'}}$ is the pairwise second order
interactions between factors $j$ and $j'$ on the output, with $1\le j<j'\le p$. These two effects were calculated by using the least squares method.
The FFD being saturated, the residual terms $E_i$ were all zero. The residual variance could not be therefore estimated. Since the NitroScape model is a deterministic model, the residual variance would have only corresponded to interactions of order higher than two. If needed,
this variance could have been estimated by using techniques based on a parsimony principle to extract some degrees of freedom \citep{dagnelie1998plans}.
}

\new{For a given output $x_{i,j}$, the main effect of each factor $j$ is:
\begin{equation}
\nonumber
 mSI_j = \sum_{k=1}^3 \#\mathcal{X}_{j}^{(k)} \cdot (\bar Y^{(k)}_j - \bar Y)^2 \bigg/ TSS \,
\end{equation}
where $\bar Y=\frac{1}{n}Y_i$ is the overall average of $Y_i$'s, $\mathcal{X}_{j}^{(k)}= \{1\le i\le n : x_{i,j}=k\}$ are the sets of configurations $i$ such that the factor $j$ has level $k$, 
$\#$ denotes the cardinal of a set,
$\bar Y^{(k)}_j = 1/\#\mathcal{X}_{j}^{(k)}\cdot \sum_{i\in\mathcal{X}_{j}^{(k)} } Y_i$ are the
means for the levels $k$ of factor $j$ and  $TSS=\sum_{i=1}^n (Y_i - \bar Y)^2$ is the total sum of squares.
\\
\\
For each $1\le j<j'\le p$, the pairwise interaction effects are given by:
\begin{equation}
\nonumber
 SI_{j,j'} =  \sum_{k,k'=1}^3 \#\mathcal{X}_{j,j'}^{(k,k')}
 (\bar Y_{j,j'}^{(k,k')} -\bar Y^{(k)}_j  -\bar Y^{(k')}_{j'}   +\bar Y )^2 \bigg/ TSS
\end{equation}
where $\mathcal{X}_{j,j'}^{(k,k')}= \{1\le i\le n : x_{i,j}=k \text{ and } x_{i,j'}=k'\}$ are the sets of configurations $i$ such that the factor 
$j$ (resp. $j'$) has level $k$ (resp. $k'$) and $\bar Y_{j,j'}^{(k,k')}=1/\#\mathcal{X}_{j,j'}^{(k,k')}\cdot \sum_{i\in\mathcal{X}_{j,j'}^{(k,k')}} Y_i$. \\ \\
We also defined for each factor $j$ an index summing pairwise interaction effects involving this factor:
\begin{equation}
\nonumber
 iSI_j= \sum_{j':j'\not=j}SI_{j,j'}
\end{equation}
an index describing the total ($i.e.$ main and interaction) effect of factor $j$:
\begin{equation}
\nonumber
 tSI_j= mSI +iSI_j
\end{equation}
and an index describing the sum of interactions between all factors:
\begin{equation}
\nonumber
i_{tot} = \sum_{1\le j <j'\le p}SI_{j j'}
\end{equation}
}

The FFD being saturated, the sum of the main effects of all factors ($mSI_j$) and of the ensemble of pairwise interactions 
($i_{tot}$) added up to $100\%$ of the total variance explored by the experimental design.
Thus, $i_{tot}$ was used as a direct measure of the variance that could not be attributed to any single factor.

\subsection{Principal Component Analysis} 

The principal component analysis (PCA) is a method to transform any set of possibly correlated variables into a set
of linearly uncorrelated variables called principal components (PC). Geometrically speaking, the PCA transforms an original data set into a new data set displayed in a new orthogonal coordinate system that is defined in such a way that the greatest variance computed after projection of the data corresponds to the first axis of the orthogonal coordinate system ($i.e.$ the first principal component). We used PCA on two different kinds of data sets.

First, PCA was applied on each aggregated output to reduce data redundancy and identify features linked to the model structure, such as seasonality in time series (Section \ref{subsec:Dynamic sensitivity analysis}) or land use attribution in maps (Section \ref{subsec:Spatial sensitivity analysis}). \new{In the case of time series,
PCA was applied to the $\mathbf{Y}$ (= $(Y_{it})_{\ 1\le i\le 243,\ 1\le t\le 36}$) data set that describes temporal outputs simulated 
at the catchment outlet or spatially-aggregated outputs. Each row of the data set corresponded to each of the 243 configurations of the FFD and each column corresponded to each of the 36 months of the three-year period of interest. In the case of maps, PCA was applied to the $\mathbf{Y}$ (= $(Y_{is})_{\ 1\le i\le 243,\ 1\le s\le nc}$) data set that describes spatially-distributed and temporally-aggregated outputs. Each row of the data set corresponded to each configuration of the FFD and each column to each of the total number of grid cells ($e.g.$ $nc$=19,600 grid cells of size 12.5 m x 12.5 m each in the case of the highest horizontal resolution).} We used the R package \textit{Multisensi} \citep{lamboni2009multivariate} to carry out this analysis. 

Second, PCA was applied to the ensemble of sensitivity indices of the ensemble of temporally- and spatially-aggregated outputs,
in order to better visualize the outputs that had similar responses to input factors and evaluate the relationship between the overall
effects of the different factors. \new{PCA was applied to the data set $\mathbf{S}$ (=($S_{ij})_{\ 1\le i\le 243,\ 1\le j\le 66}$),
in which each row corresponds to each of the 243 configurations of the FFD and each column corresponds to each of the 11 main sensitivity indices 
and each of the 55 ($= \binom{11}{2}$) pairwise interaction indices.} 
We used the R package FactoMineR \citep{le2008factominer} to carry out this analysis.

\subsection{Cluster analysis}

While PCA was used to provide a reduced data set of attributes that describes the main trends in original data sets, clustering is a method we used to define groups of similar objects, based on their attribute values \citep{kaufman2009finding}. We used clustering methods with two different purposes.  

First, for each output variable, the 243 time series simulated from the different configurations of the FFD were split into three clusters that grouped curves with similar features ($e.g.$ slope, range of variation). \new{This clustering was performed by using the R package \textit{KML} \citep{kml} that is based on a k-means algorithm \citep{steinhaus1956division} applied to the features of the curves.}
\new{The number of clusters was set to three which corresponds to the number of levels for each input factor in the FFD.}
\new{This clustering was a first approach to visualize the separation between time series and to detect on which feature they might differ. The obtained clusters of time series were represented using the same method as that described in Subsection \ref{ssec:visualization}.}

Second, \new{a hierarchical clustering \citep{ward1963hierarchical}} was applied on the ensemble of results of the sensitivity analyses of all temporally- and spatially-aggregated outputs \new{($i.e.$ the $\mathbf{S}$ data set described in Section 2.5)}, in order to synthesize the results obtained for the ensemble of outputs. The R packages \textit{FactoMineR} and \textit{PVclust} \citep{suzuki2006pvclust} were used to carry out this clustering.
The joint application of cluster analysis and PCA provided representations that make it possible to identify groups of outputs with similar profiles of sensitivity indices and better visualize the relations between the effects of input factors on the ensemble of outputs. Such representations led to three kinds of interpretation. First, orthogonality between the projections of the sensitivity indices of two factors indicated \new{that the effects of the two factors were independent:} outputs might be affected by either one factor, both of them or any of them. Second, the parallel projection of the sensitivity indices of two factors indicated that whenever one of the factors had an effect on a given output, the other factor had an effect too. Third, the antiparallel projection of the sensitivity indices of two factors indicated that whenever one of the factors had an effect on a given output, the other factor did not have any effect and \textit{vice versa}.

\section{Results and Discussion}

This section shows and discusses a few examples of the detailed sensitivity analysis applied on the 29 $N_r$-related output variables of the NitroScape model. Section \ref{subsec:Dynamic sensitivity analysis} compares the temporal sensitivity analysis of two spatially-aggregated variables. Section \ref{subsec:Spatial sensitivity analysis} compares the spatial sensitivity analysis of two temporally-aggregated variables. The correspondence between spatial and temporal sensitivity analyses is briefly discussed in Section \ref{subsec:Correspondence}.
The results of the sensitivity analysis of the ensemble of the 29 spatially- and temporally-aggregated outputs are summarized in Section \ref{subsec:Synthese}. Extracting conclusions from the ensemble of results of the detailed spatial and temporal sensitivity analyses is out of the scope of this study.

\subsection{Temporal sensitivity analysis}
\label{subsec:Dynamic sensitivity analysis}

A temporal sensitivity analysis was applied on each spatially-aggregated output and on each output describing the catchment outlet. Figure \ref{fig:AS_CL_dyn_NOx} (resp. Fig. \ref{fig:AS_CL_dyn_NH4up}) outlines the detailed results of the temporal sensitivity analysis performed on two examples of $N_r$ fluxes between landscape compartments: $NO_x$ emissions from agroecosystems to the air (resp. soil $NH_4^+$ uptake by plants), cumulated from the beginning of the three-year period of interest and for the whole landscape.

Some remarks can be extracted from Figures \ref{fig:AS_CL_dyn_NOx} and \ref{fig:AS_CL_dyn_NH4up}:

\begin{figure}
  \centering
  \includegraphics[width=\linewidth]{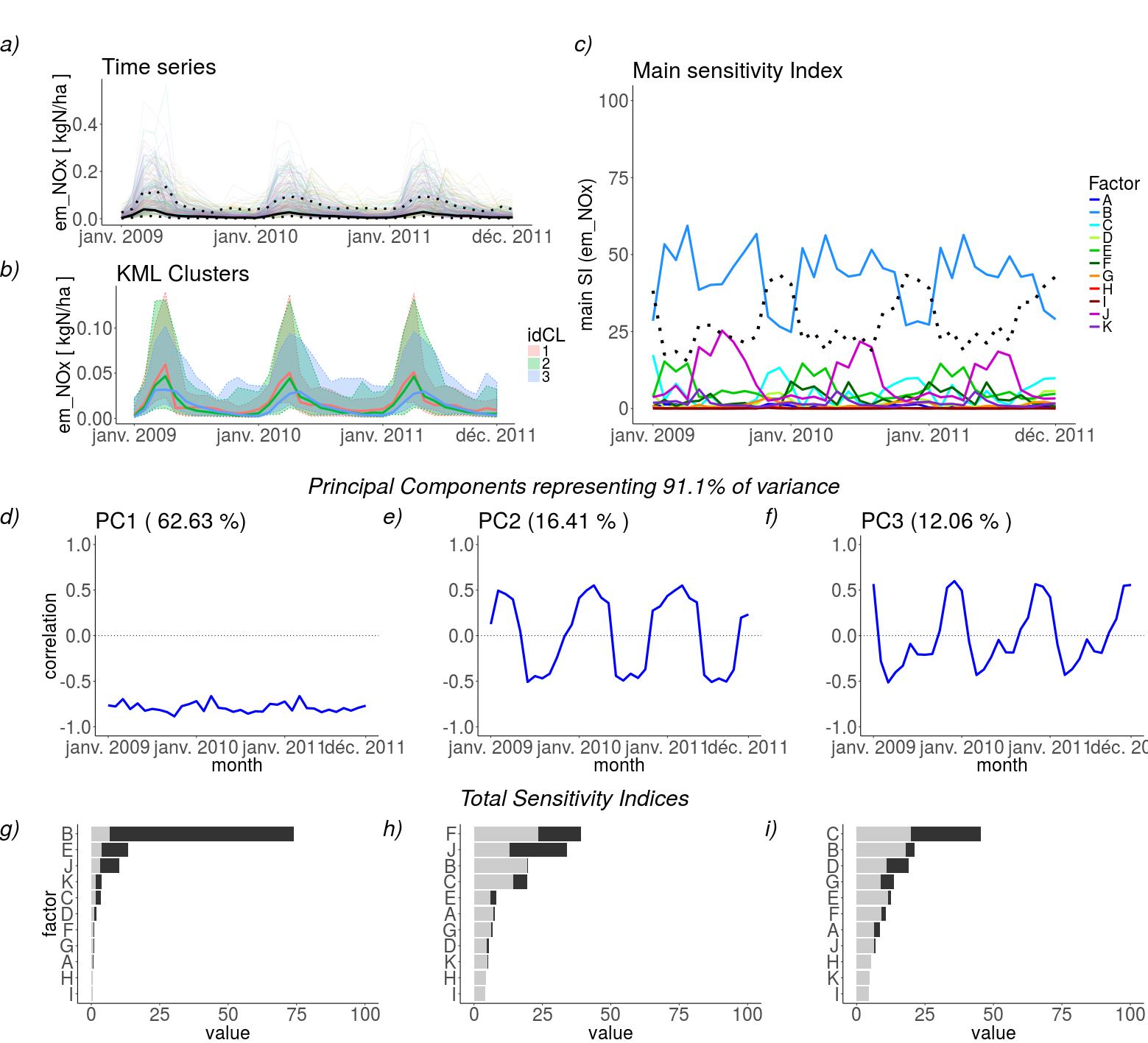}
  \caption{Temporal sensitivity analysis of $NO_x$ emissions simulated for the whole landscape and averaged by area unit; (a) time series of each simulated configuration of the numerical experiment (colored lines), central time series (bold black line) and middle region (dashed black line);
  (b) time series of three clusters grouping most-similar curves \new{via a k-means algorithm for longitudinal data (KML)}, idCL is cluster label;
  (c) temporal main sensitivity indices of each factor (colored lines) and of the sum of interactions between factors (dashed black line).
  \new{Sensitivity analysis on each PC}: (d,e,f) decomposition of the first three principal components (PC); (g,h,i) total sensitivity indices of each factor on each PC, split into main (black bars) and pairwise interaction (gray bars) effects.}
  \label{fig:AS_CL_dyn_NOx}
\end{figure}

\begin{figure}
  \centering
  \includegraphics[width=\linewidth]{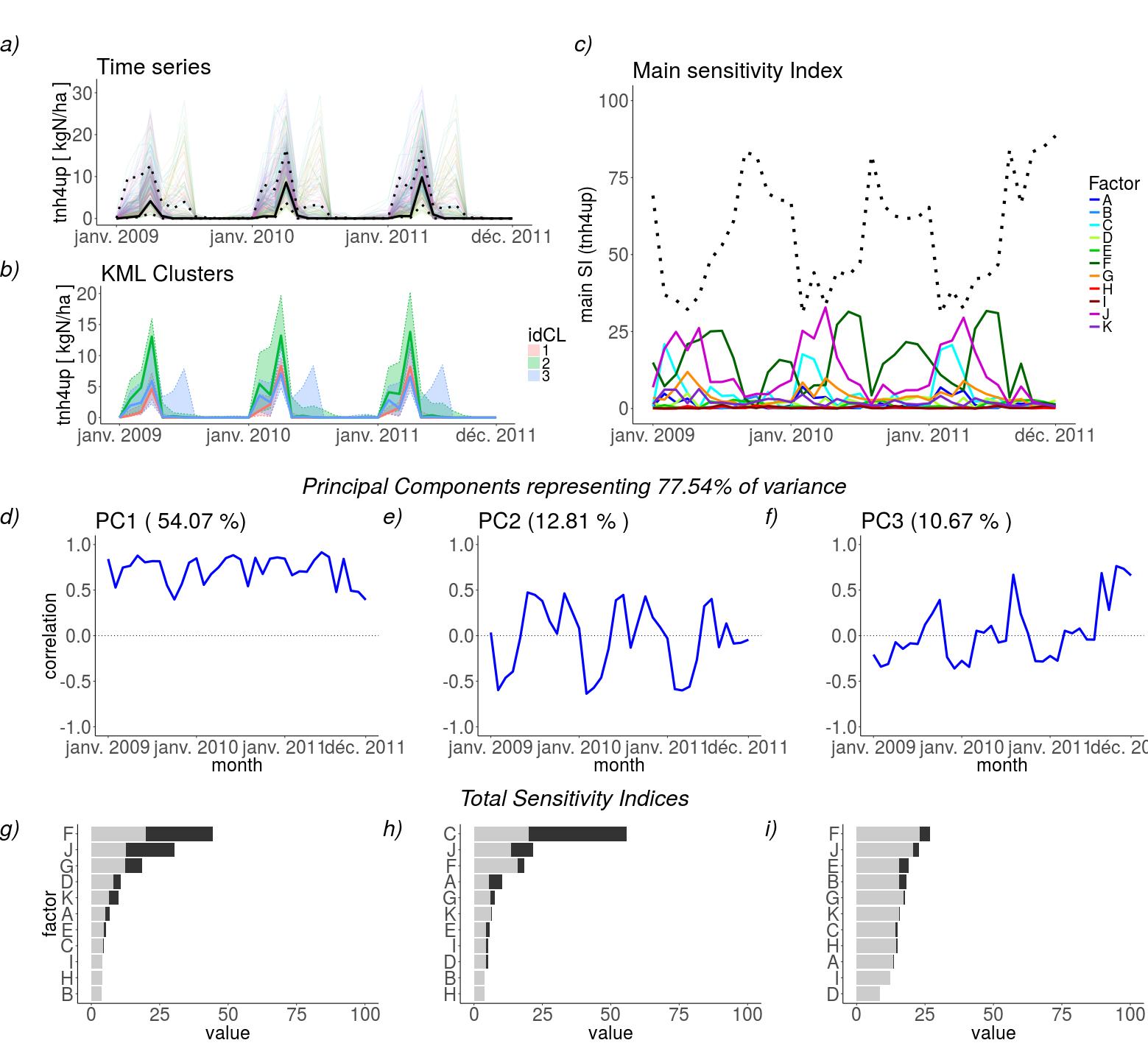}
  \caption{Temporal sensitivity analysis of $NH_4^+$ uptake by plants simulated for the whole landscape and averaged by area unit; (a) time series of each simulated configuration of the numerical experiment (colored lines), central time series (bold black line) and middle region (dashed black line);
  (b) time series of three clusters grouping most-similar curves, idCL is cluster label;
  (c) temporal main sensitivity indices of each factor (colored lines) and of the sum of interactions (dashed black line).
  \new{Sensitivity analysis on each PC}: (d,e,f) decomposition of the first three principal components (PC); (g,h,i) total sensitivity indices of each factor on each PC, split into main (black bars) and pairwise interaction (gray bars) effects.}
  \label{fig:AS_CL_dyn_NH4up}
\end{figure}

\begin{enumerate}[i]
 \item Time series showed peaks of both $NO_x$ emissions and $NH_4^+$ uptake during spring fertilization periods \new{(Fig. \ref{fig:AS_CL_dyn_NOx}a and \ref{fig:AS_CL_dyn_NH4up}a)}.
 \item Clusters grouped time series based on their mean over time, range of peaks and dynamic variance. 
 \new{The clustered time series for $NH_4^+$ uptake were quite different since the averaged time series by cluster were well separated and the time series in cluster 3 had a second peak in summer while the three clusters of time series for $NO_x$ emissions nearly overlapped
 }\new{(Fig. \ref{fig:AS_CL_dyn_NOx}b and \ref{fig:AS_CL_dyn_NH4up}b)}.
 \item $NO_x$ emissions were mostly sensitive to the vertical resolution of the model (factor B: $mSI_B = (41 \pm 7 )\%$) 
 and to the sum of pairwise interactions ($i_{tot} = (28 \pm 6 )\%$). $NH_4^+$ uptake was mostly affected by pairwise interactions of multiple factors ($i_{tot} = (56 \pm 18 )\%$). $NH_4^+$ uptake was also sensitive to the main effects of soil surface porosity (factor F), fertilization type (factor J) and soil lateral transmissivity (factor C)
 \new{(Fig. \ref{fig:AS_CL_dyn_NOx}c and \ref{fig:AS_CL_dyn_NH4up}c)}.

 \item  PC1 represented \mnew{roughly the average of the time-series (Fig. \ref{fig:AS_CL_dyn_NOx}d and \ref{fig:AS_CL_dyn_NH4up}d).} 
 For $NO_x$ emissions, PC1 was mainly sensitive to the main effects of vertical resolution (factor B), while variations in $NH_4^+$ uptake came mosly from pairwise interactions involving soil surface porosity and fertilization type (factors F and J). \new{This result is consistent with the large gray bars shown on Fig. \ref{fig:AS_CL_dyn_NOx}g and \ref{fig:AS_CL_dyn_NH4up}g that represent pairwise interactions}.
 
 \item  PC2 revealed the factors that mostly affect time series with one-year periodicity ($e.g.$ the factors that mostly affect time series during spring). For $NO_x$ emissions, PC2 mainly reflected the effects of soil surface porosity (factor F), fertilization type (factor J) and their pairwise interactions. For $NH_4^+$ uptake, PC2 mainly depended on the main effect of soil lateral transmissivity (factor C) \new{(Fig. \ref{fig:AS_CL_dyn_NOx}e and \ref{fig:AS_CL_dyn_NOx}h, Fig. \ref{fig:AS_CL_dyn_NH4up}e and \ref{fig:AS_CL_dyn_NH4up}h)} and its pairwise interactions .  
 \item PC3 captured effects with smaller seasonality, showing peaks of representation on the zeros of PC2. For $NO_x$ emissions, PC3 captured the effects of soil lateral transmissivity (factor C). For $NH_4^+$ uptake, PC3 mainly captured the ensemble of pairwise interactions
 \new{(Fig. \ref{fig:AS_CL_dyn_NOx}f and \ref{fig:AS_CL_dyn_NOx}i, Fig. \ref{fig:AS_CL_dyn_NH4up}f and \ref{fig:AS_CL_dyn_NH4up}i)}.
\end{enumerate}


\subsection{Spatial sensitivity analysis}
\label{subsec:Spatial sensitivity analysis}

\begin{figure}
  \centering
  \includegraphics[width=\linewidth]{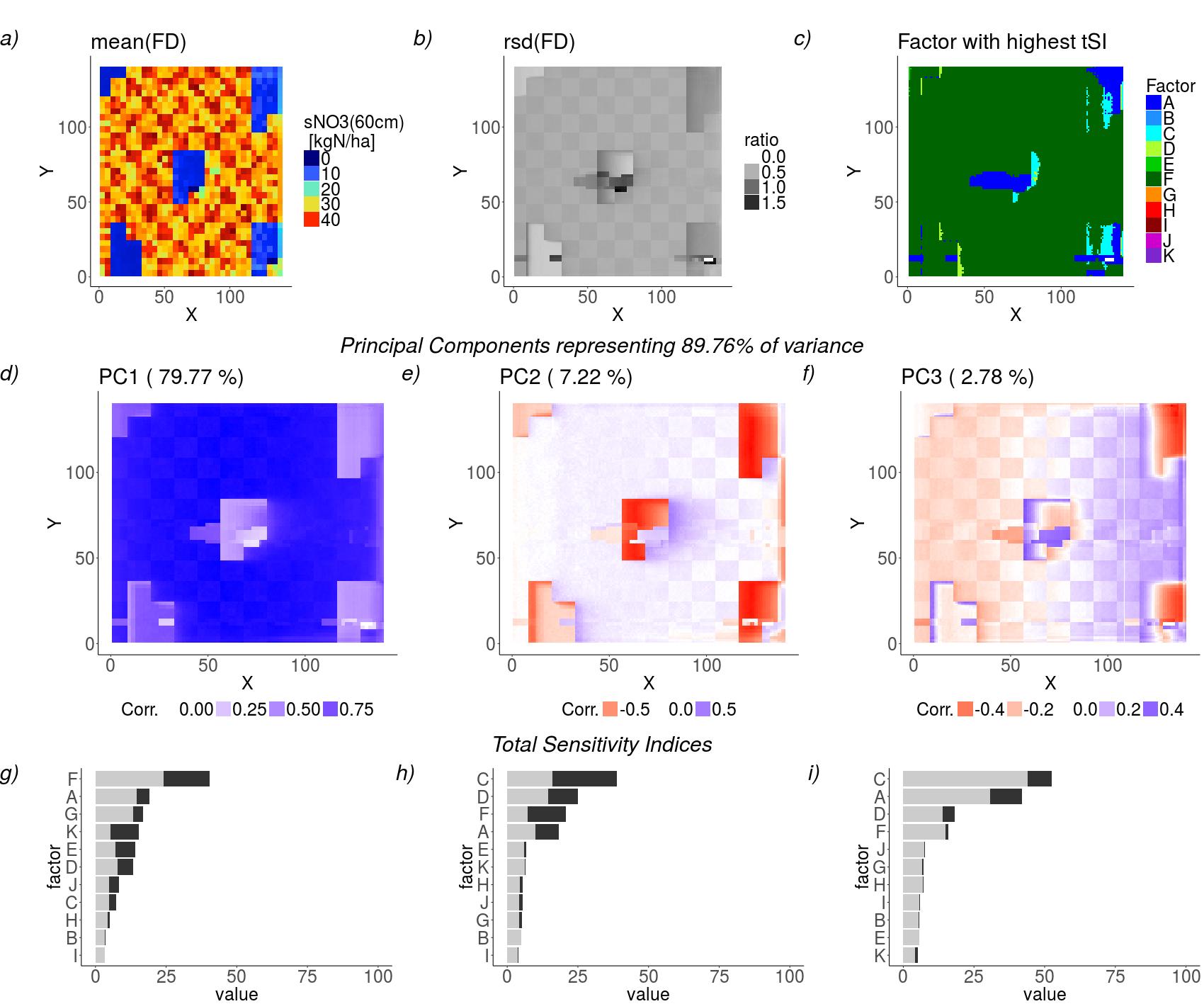}
  \caption{Spatial sensitivity analysis of soil $NO_3^-$ amount between 0 and 60 cm depth cumulated on the three-year period of interest in each grid cell of the landscape and averaged by area unit; (a) central map of averages over time within the fractional factorial design (FFD); (b) rsd: coefficient of variation between configurations of the FFD
  \new{averaged over time};
(c) map of the factors with the highest total sensitivity index (tSI) in each grid cell. \new{Sensitivity analysis on principal components}: (d,e,f) decomposition of the first three principal components; (g,h,i) total sensitivity indices of each factor on each PC, split into main (black bars) and interaction (gray bars) effects.}
  \label{fig:AS_2Dmap_sno3}
\end{figure}

\begin{figure}
  \centering
  \includegraphics[width=\linewidth]{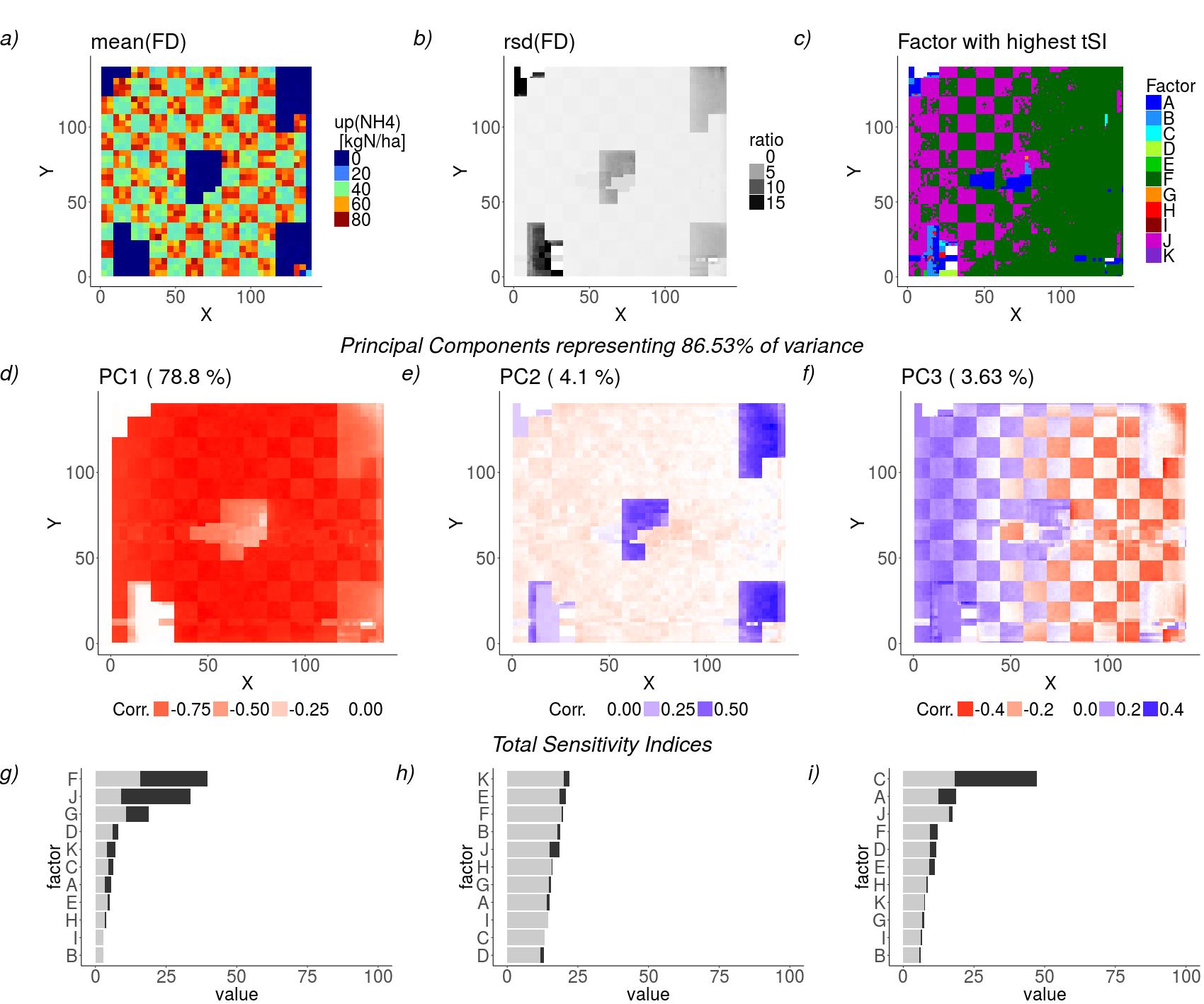}
  \caption{Spatial sensitivity analysis of $NH_4^+$ uptake by plants cumulated on the three-year period of interest in each grid cell of the landscape and averaged by area unit; (a) central map of averages over time within the fractional factorial design (FFD); (b) rsd: coefficient of variation between configurations of the FFD
  \new{averaged over time};
(c) map of the factors with the highest total sensitivity index (tSI) in each grid cell. \new{Sensitivity analysis on principal components}: (d,e,f) decomposition of the first three principal components; (g,h,i) total sensitivity indices of each factor on each PC, split into main (black bars) and interaction (gray bars) effects.}
  \label{fig:AS_2Dmap_NH4up}
\end{figure}

\new{Figure \ref{fig:AS_2Dmap_sno3} (resp. Fig. \ref{fig:AS_2Dmap_NH4up}) outlines the results of spatial sensitivity analysis for the amount of soil $NO_3^-$ between 0 and 60 cm depth (resp. the amount of soil $NH_4^+$ uptake by plants) in each grid cell for the three-year period of interest.} 

Some remarks can be extracted from Figures \ref{fig:AS_2Dmap_sno3} and \ref{fig:AS_2Dmap_NH4up}:
\begin{enumerate}[i]
 \item Both soil $NO_3^-$ amount and $NH_4^+$ uptake by plants were smaller for unmanaged grasslands than for croplands \new{(Figs \ref{fig:AS_2Dmap_sno3}a and \ref{fig:AS_2Dmap_NH4up}a)}.
 
 \item Conversely \new{for $NH_4^+$ uptake}, the relative variance was greater in unmanaged grasslands and around farm buildings, indicating that these areas were more sensitive to input factors
 \new{(Figs \ref{fig:AS_2Dmap_sno3}b and \ref{fig:AS_2Dmap_NH4up}b)}.

 \item  For both variables, the factors with the highest effect were spatially distributed. The effect of horizontal resolution ($i.e.$ grid cell width, factor A) was the highest around farm buildings and on the edges of the landscape. Elsewhere, the factors having the highest effects varied throughout the landscape depending on elevation and land use. 
 Generally, the soil surface porosity (factor F) was the factor having the highest effect on both variables, although the type of fertilization (factor J) was the paramount factor for maize crops located upslope \new{for $NH_4^+$ uptake} \new{(Fig. \ref{fig:AS_2Dmap_sno3}c and \ref{fig:AS_2Dmap_NH4up}c)}.
 \item For both variables, PC1 described roughly the spatial mean of FFD variance. PC1 was mostly sensitive to the main effect of soil surface porosity (factor F) and to its pairwise interactions 
 \new{(Fig. \ref{fig:AS_2Dmap_sno3}d and \ref{fig:AS_2Dmap_sno3}g, Fig. \ref{fig:AS_2Dmap_NH4up}d and \ref{fig:AS_2Dmap_NH4up}g)}.
 
 \item For both variables, PC2 was strongly correlated with unmanaged grasslands downslope and less correlated with croplands and upslope areas. For soil $NO_3^-$ amount, PC2 was mostly sensitive to soil surface porosity (factor F), while for $NH_4^+$ uptake, PC2 was mostly affected by pairwise interactions
 \new{(Fig. \ref{fig:AS_2Dmap_sno3}e and \ref{fig:AS_2Dmap_sno3}h, Fig. \ref{fig:AS_2Dmap_NH4up}e and \ref{fig:AS_2Dmap_NH4up}h)}.
 
 \item  For both variables, PC3 exhibited more complex correlations with the landscape slope and the checkerboard distribution of croplands. For soil $NO_3^-$ amount, PC3 was mostly affected by pairwise interactions, while for $NH_4^+$ uptake, PC3 was mainly affected by soil lateral transmissivity
 \new{(factor C, Fig. \ref{fig:AS_2Dmap_sno3}f and \ref{fig:AS_2Dmap_sno3}i, Fig. \ref{fig:AS_2Dmap_NH4up}f and \ref{fig:AS_2Dmap_NH4up}i)}.
 
\end{enumerate}


\subsection{Correspondence between spatially explicit and temporal sensitivity analyses}
\label{subsec:Correspondence}

Figures \ref{fig:AS_CL_dyn_NH4up} and \ref{fig:AS_2Dmap_NH4up} represent two different aspects of the detailed sensitivity analysis of the cumulated $NH_4^+$ uptake by plants. The joint analyses of the spatially-agregated and temporally-agregated data sets made it possible to analyse the effects of the input factors on output variables from two complementary points of view and offered a more comprehensive visualization of the effects of the input factors. For instance, the time series (Fig. \ref{fig:AS_CL_dyn_NH4up}) show that during fertilization periods, the paramount factor throughout the whole landscape was the type of fertilizer (factor J). Soil surface porosity (factor F) also played a predominant role \new{just after fertilization periods}. In parallel, the spatial map (Fig. \ref{fig:AS_2Dmap_NH4up}) shows that fertilizer type had a greater effect upslope and the effect of soil surface porosity was greater downslope. Such a joint analysis indicate that the amount of $NH_4^+$ uptaken by plants was highly dependent on the percolation dynamics of the fertilizer.     

A detailed analysis of how each factor affects each variable is out of the scope of this study. 

\subsection{Classification of the outputs regarding the sensitivity indices}
\label{subsec:Synthese}

\new{A cluster analysis was applied to the $29$ temporally- and spatially-aggregated outputs on the basis of their sensitivity indices.
This led to groups of outputs having similar response to input factors.}


\begin{figure}
  \centering
  \includegraphics[width=\linewidth]{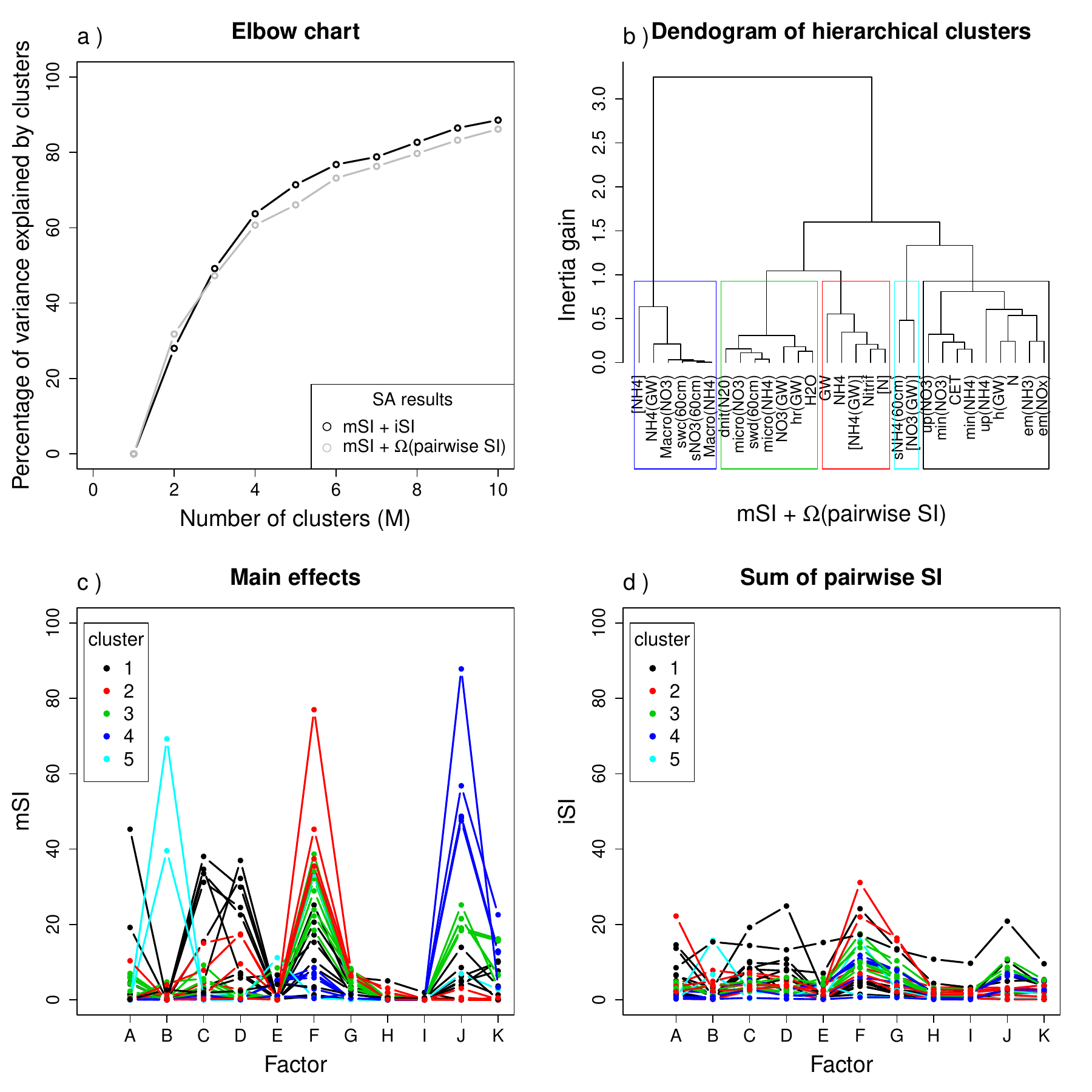}
  \caption{Cluster analysis of the \new{29} temporally- and spatially-agregated outputs based on their sensitivity index profiles; (a) percentage of variance explained by clusters as a function of the number M of clusters; SA results are expressed either in terms of the main effects (mSI) and the sum of pairwise interactions (iSI) of each factor (black line), or in terms of the main effects of each factor (mSI) and the ensemble of pairwise interactions $\Omega$(pairwise SI) (gray line); (b) hierarchical clustering of outputs in which outputs are linked together if they have similar profiles of sensitivity indices; Inertia gain (y-axis) is the variance explained when outputs are linked together. Color boxes indicate the clusters obtained for M = 5; (c) main effects of each factor on each output; (d) sum of pairwise interactions of each factor on each output. Colors of each line are set according to the colors of clusters.}
  \label{fig:SI_Clusters}
\end{figure}

Figure \ref{fig:SI_Clusters} shows the clusters into which model outputs were split. The number of clusters (M = 5) was set as the minimal number providing equal classifications of the outputs with different clustering algorithms (k-means and hierarchical clustering). This partitioning made it possible to explain $73.6\%$ of the variance of the sensitivity indices and the number of clusters found by this way corresponded to the number that would be chosen qualitatively with the elbow method  \citep{ketchen1996application}.

\begin{figure}
  \centering
 {\includegraphics[scale=.35]{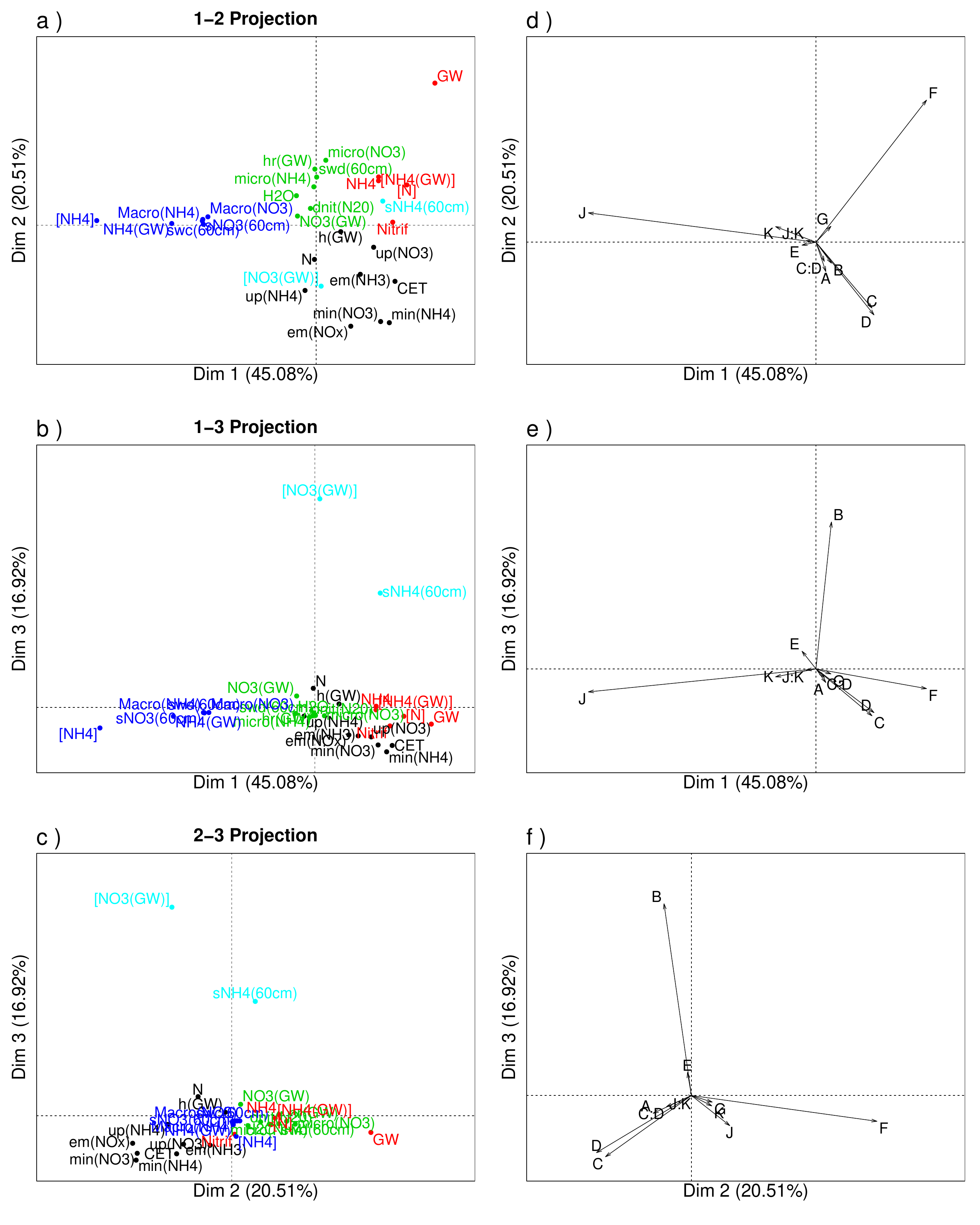}}
  \caption{Principal component analysis and clustering of the results of the sensitivity indices resulting from the analysis of \new{the 29} temporally- and spatially-agregated outputs; (a,b,c) projections of the clusters of outputs onto the plane defined by two principal components; (d,e,f) projections of sensitivity indices of input factors onto the same planes.
  \new{Clusters are identified by their color which are the same as in Figure 6.}}
  \label{fig:resSI_PCA}
\end{figure}

The principal projections of the clusters of outputs onto the axes of the transformed space are shown in Figures \ref{fig:resSI_PCA}a, \ref{fig:resSI_PCA}b and \ref{fig:resSI_PCA}c. The corresponding principal projections of the sensitivity indices of input factors onto the axes of the transformed space are shown in Figures \ref{fig:resSI_PCA}d, \ref{fig:resSI_PCA}e and \ref{fig:resSI_PCA}f.

The PC1-PC2 projection explained $65.6\%$ of the variance of the sensitivity indices. This projection made it possible to clearly discriminate clusters 1, 2 and 4, but clusters 3 and 5 were not so easily singled out (Fig. \ref{fig:resSI_PCA}a). This observed cluster separation was driven by the main effects of factors J, F and also factors C and D to a lesser extent. Indeed, clusters were split along the axes indicated by the arrows corresponding to the main effect of these factors, the length of each arrow being proportional to the importance of each effect (Fig. \ref{fig:resSI_PCA}d). In such a projection, orthogonality indicates that indices are independent from each other: the effects of factors J and F were almost independent from each other, as well as the effects of factors F, C and D. In contrast, factor J was antiparallel to factors C and D, indicating that whenever factor J had an effect, the other two did not, and \textit{vice versa}. Finally, factors C and D were parallel to each other indicating that they had the same effect on the same clusters of variables. 
  
The PC1-PC3 projection explained $62\%$ of the variance. In such a projection, cluster splitting was driven by the main effects of factors J, B and F (Fig. \ref{fig:resSI_PCA}b and \ref{fig:resSI_PCA}e). Cluster 5 was separated along the axis of the main effect of factor B, while cluster 2 could be singled out along the axis of the main effect of factor F, and in opposition to the axis of the main effect of factor J. That means that variables in cluster 2 were affected by the main effect of factor F and not by the main effect of factor J.

The PC2-PC3 projection explained $37.5\%$ of the variance (Figures \ref{fig:resSI_PCA}c and \ref{fig:resSI_PCA}f). It made it possible to clearly discriminate cluster 5 as well as splitting the other clusters along the axes of the main effects of factors C, D and F. 

Figure \ref{fig:resSI_PCA} shows that the total amount of $NO_3^-$ discharge at the catchment outlet is always located near the origin of the coordinate system. That indicates that this variable was equally affected by the main effects and pairwuise interactions of each factor appearing in the projections.   

Table \ref{tab : synthese_SI} summarizes the results of the cluster analysis and the PCA applied on the ensemble of spatially- and temporally- aggregated outputs, characterized by their sensitivity indices. 

 \begin{table}[ht!]
     \centering
     \resizebox{\textwidth}{!}{%
  \begin{tabular}{|l|c|l|l|}%
    \hline
    Cluster&N &Output variables&Characteristics \\
    \hline
    K=1 / black&9& \multicolumn{1}{p{6cm}|} {Evapotranspiration, nitrogen emission and uptake by plants, nitrogen mineralization, depth of the groundwater table and total amount of $NO_3^-$ at the catchment outlet.} &  \multicolumn{1}{p{9cm}|} {Mostly affected by soil lateral water transmissivity (factor C) and soil transmissivity decrease with depth (factor D).} \\
    K=2 / red &5& \multicolumn{1}{p{6cm}|} {Total amount of $NO_3^-$ and $NH_4^+$ discharged at the catchment outlet, nitrification, total amount of $NO_3^-$ and $NH_4^+$ in groundwater.} & \multicolumn{1}{p{9cm}|} {Mostly affected by soil surface porosity (factor F).}\\
    K=3 / green&7& \multicolumn{1}{p{6cm}|} {Surface water depths, streaming flow and water discharge at the catchment outlet, nitrogen adsorbed in soil microporosity, soil $NO_3^-$ in groundwater.} & \multicolumn{1}{p{9cm}|} {Mostly affected by soil surface porosity (factor F), fertilization type (factor J), soil lateral transmissivity (factor C) and soil transmissivity decrease with depth (factor D). Moderate effect of interaction terms.} \\
    K=4 / dark blue&6& \multicolumn{1}{p{6cm}|} {$NH_4^+$ concentration in groundwater and at the catchment outlet, nitrogen adsorbed in soil macroporosity, soil $NO_3^-$ concentration in soil surface.} & \multicolumn{1}{p{9cm}|} {Mostly affected by fertilization type (factor J), high effect of the interaction term J:K.}\\
    K=5 / light blue&2& \multicolumn{1}{p{6cm}|} {$NH_4^+$ concentration in soil surface and $NO_3^-$ concentration in groundwater.}& \multicolumn{1}{p{9cm}|} {Mostly affected by vertical resolution (factor B).}\\
    \hline
  \end{tabular}}
  \caption{\new{Description of the five clusters of spatially- and temporally-aggregated outputs found by a hierarchical clustering on their corresponding main and second order sensitivity indices.
 The cluster columns provides the number of each cluster and its related color on Figures 6 and 7. The $N$ column provides the number of output variables included in each cluster.}} 
  \label{tab : synthese_SI}
\end{table}


Clusters grouped variables that were sensitive to the same factors. However, this did not entail that these factors affected those variables in the same way: for instance, soil $NO_3$ amount between 0 and 60 cm depth ($sNO_3(60cm)$) and $NH_4$ concentration in groundwater ($NH_4(GW)$) were grouped together in cluster 4 as they both had a high sensitivity to soil lateral transmissivity (factor C), while $sNO_3(60cm)$ decreased and $NH_4(GW)$ increased when the level of factor C increased.

In broad terms, model outputs were mostly affected by the hydrological characteristics of soil and management ($i.e.$ fertilization type). Interaction terms had significant effects on the detailed sensitivity analyses of every output, but they were less important for spatially- and temporally-aggregated output variables.    
The model resolution did have a significant effect on some model outputs, comparable to the effect of other input factors. The horizontal resolution of the model (A) had a significant effect on several variables, but only for some areas of the landscape and not at the aggregated level.
The vertical resolution (B) had a significant effect on the two spatially- and temporally-aggregated variables soil $NH_4^+$ amount between 0 and 60 cm depth and $NO_3^-$ concentration in groundwater. 

\section{Conclusions}


We developed a framework to perform a thorough and comprehensive sensitivity analysis of a complex model with numerous scalar input factors and multiple spatially distributed and temporal output variables. We implemented methods for computing various statistical indicators, visualizing and aggregating the model outputs, and synthesizing the ensemble of results of the sensitivity analyses. 

The synthesis of results made it possible to classify output variables according their responses to the ensemble of input factors, as well as to classify input factors according to their effect on the ensemble of outputs. In particular, our methods indicated that spatial resolution did have an effect on model behavior \new{since sensitivity indices of factors A and B were found to be large for several output variables}.
The presented methods could be used for reducing the dimensionality of the space of input factors, because they make it possible to rule out factors that have nearly no effect on the outputs within the range of the explored values and in this particular theoretical landscape, such as the ratio of soil microporosity to macroporosity (factor G) or the depth of the intermediate soil layer (factor H). \new{For the most influential factors, the sensitivity analysis could be refined by using other indicators such as Sobol indices \citep{saltelli2000sensitivity,gamboa2014sensitivity} or HSIC \citep{da2015global,de2017sensitivity}.
}

The detailed analysis of sensitivity of every output variables to the numerous input factors was made possible by aggregating either spatially or temporally the output variables. Other types of data aggregation could be applied: for instance, data could be aggregated by land use, $e.g.$ by grouping together all grid cells with the same land use. Output variables could also be aggregated according to meteorological inputs, $e.g.$ by grouping together the days following immediately a rain event. 
\new{The scale of aggregation was shown to be an important issue since \citet{saint2012,saint2014} reported that the most influent input factors were not the same at different scales of aggregation.}
Such aggregations could be used to compare different types of agricultural management strategies or to design alternative managing responses to meteorology,
\new{e.g. to determine physical features and agricultural management strategies that mostly affect output variables after rain events or on grasslands.}




\section*{Acknowledgements}

This work was supported by the French Research Agency (ANR), Agrobiosphere program, ESCAPADE project (ANR-12-AGRO-0003).

\end{document}